\documentclass[aps,prd,amssymb,eqsecnum]{revtex4}

\usepackage{graphicx}
\usepackage{bm}

\begin{document}


%
%
{\baselineskip0pt
\rightline{\large\baselineskip14pt\rm\vbox
        to20pt{\hbox{OU-TAP-230}
	       \hbox{IGPC-04/5-2}
               \hbox{KUNS-1916}
\vss}}
}
\vskip15mm

\title{
Gravitational Wave Memory of Gamma-Ray Burst Jets
}

\author{Norichika Sago$^1$, Kunihito Ioka$^2$,
Takashi Nakamura$^3$ and Ryo Yamazaki$^1$}
\affiliation{$^1$Department of Earth and Space Science,
Graduate School of Science, Osaka University, Toyonaka 560-0043, Japan
\\
$^2$Physics Department, Pennsylvania State University,
University Park, PA 16802
\\
$^3$Department of Physics, Kyoto University,
Kyoto 606-8502, Japan
}


\begin{abstract}
Gamma-Ray Bursts (GRBs) are now considered as relativistic jets.
We analyze the gravitational waves from the acceleration stage of the GRB jets.
We show that (i) the point mass approximation is not appropriate 
if the opening half-angle of the jet is larger than
the inverse of the Lorentz factor of the jet,
(ii) the gravitational waveform  has many step function like jumps, and
 (iii) the practical DECIGO and BBO may detect
such an event if the GRBs occur in the Local Group.
We found that the light curve of GRBs and the gravitational waveform 
are anti-correlated so that the detection of the gravitational wave
is indispensable to determine the structure of GRB jets.
\end{abstract}

\pacs{Valid PACS appear here}

\maketitle

\section{Introduction}

The gravitational wave astronomy enters its first step.
The ground-based laser interferometers such as TAMA300, LIGO and GEO600
have already begun operations covering $10$ Hz -- kHz band,
and VIRGO will be also in operation soon.
The Laser Interferometer Space Antenna (LISA) \cite{LISA}
covering $10^{-4}$ -- $10^{-2}$ Hz is expected to be launched in $2011$.
Recently, in addition, the Decihertz Interferometer Gravitational wave
Observatory (DECIGO), which aims the frequency band around decihertz
($10^{-2}$ -- $10$ Hz), has been proposed \cite{Seto:2001qf}.
The decihertz antenna is also considered in the NASA Roadmap Beyond Einstein
and is called the Big Bang Observer (BBO) \cite{BBO}.

It might be possible to construct a practical DECIGO and BBO 
with $h_{\rm rms}\sim 3\times 10^{-24}$ at $\sim 0.1$ Hz.
Although this limiting sensitivity is $\sim 10^{3}$ times less than
the ultimate quantum limit $h_{\rm rms}\sim 3 \times 10^{-27}$
at $\sim 0.1$ Hz for a $100$ kg mirror mass, the practical DECIGO and BBO
may achieve the following important results
\cite{Seto:2001qf,Takahashi:2003wm}:
(1) The chirp signals of coalescing binary neutron stars
at $z\sim 1$ may be detected with a signal-to-noise ratio (SNR) $\sim10$
for one year observation.
(2) The time variation of the Hubble parameters of our universe 
may be determined by the chirp signals so that we can directly measure
the acceleration of the expansion of the universe.
(3) The stochastic gravitational waves from inflation may be detected if 
$\Omega_{\rm GW} \agt 10^{-15}$, which is the upper limit from cosmic 
microwave background.
(4) An arcminute position and an accurate ($\sim 0.1$ s) coalescence
time of the coalescing binary neutron stars
at $\sim 300$ Mpc may be determined a week before the final merging,
so that the all band electromagnetic detectors can perform the
simultaneous follow-up observation.
An accurate position may be also important for the reconstruction of the 
density distribution and the shape of the galactic dark halo
\cite{Ioka:1999bq,Ioka:1998gf,Ioka:1998nz}.

In this paper, we consider gamma-ray bursts (GRBs) 
as possible sources of the gravitational waves for DECIGO and BBO.
The GRBs are the most luminous objects in the universe.
The recent discovery of the afterglow has revolutionized
the understanding of the long duration GRBs
(e.g. \cite{Meszaros:2001vi,Zhang:2003uk}).
We are now confident that long GRBs with $\sim $10s duration eject
relativistic flows with Lorentz factors $\gamma \ge 100$ and the
ejecta is collimated in a jet.
This means that the nonspherical acceleration of a large mass 
is associated with GRBs.
Therefore it is guaranteed that the GRB jet acceleration
emits gravitational waves
\cite{Segalis:2001ns,Piran:2001ks,Piran:2002kw}.
We will show that the typical frequency of the gravitational waves is about 
the inverse of the  duration of GRBs  $\sim 0.1$ Hz,
so that the decihertz band of DECIGO and BBO best suits the detection
of gravitational waves from the GRB jet (\S~\ref{sec:light}).

Here we concentrate on the gravitational waves from the  acceleration
stage of GRB jets, while we also expect the gravitational waves 
from the formation of the  central engine of GRBs
\cite{Kobayashi:2002md,Kobayashi:2002by,vanPutten:2003hd,Fryer:2000my}
that could be detected by the ground-based laser interferometers.
We know that the accelerated jet exists although we do not know
how it is accelerated and what structure it has.
Since the jet is accelerated near the central engine
where the optical depth is quite large,
the electromagnetic probe is useless.
The high-energy neutrinos from GRBs (\cite{Levinson:2003je})
may provide a indirect probe of
the central engine, while the gravitational waves
may bring us the direct information of it.
Thus the gravitational waves could be an ultimate way
to see the acceleration regime.

The gravitational waves from the GRB jet have
a ``memory'', that is,
the metric perturbation does not return to its original value
at the end of the jet acceleration \cite{Braginsky}.
The gravitational wave memory is the change
in the transverse-traceless (TT) part of
the Coulomb-type ($\propto 1/r$) gravitational field 
generated by the masses of the system.
These masses could be stars, neutrinos
\cite{Epstein:dv,Turner:1978},
gravitons \cite{Christodoulou:cr,Thorne:1992,Wiseman:ss},
and in our case the GRB jets.

Segalis and Ori \cite{Segalis:2001ns} analyzed the gravitational wave
emitted when
an ultrarelativistic blob is ejected from a massive object
by using a point mass approximation.
They showed that the gravitational wave emission from
such a source is anti-beamed.
This contrasts with the electromagnetic emission, which is beamed
into the forward direction.
They also obtained the detectable distance of about 15 Mpc and
estimated the detection rate of about 1 per year
by the advanced LIGO, assuming that the total energy of the GRBs is
about $\sim 10^{52}$ erg and that GRBs follow supernovae.

On the other hand, recent studies suggest that the jet structure
is essentially important for understanding the GRB phenomena
\cite{Rossi:2001pk,Zhang:2001qt,Zhang:2003zg}.
Depending on the viewing angle,
the structured jet model might be able to explain the diverse
phenomena of GRBs, such as
short GRBs, long GRBs, X-ray flashes (XRFs) and X-ray rich GRBs
\cite{Yamazaki:2004ha}.
Here XRFs and X-ray rich GRBs are fast X-ray
transients that are considered to be related to long GRBs
\cite{Heise:2001,Sakamoto:2004}.
Little is  known about  short GRBs since no afterglow was observed
although many suggestions to show the similarity to long GRBs exist.

It is difficult to determine the jet structure only by
the electromagnetic waves, since the photons are beamed into
the forward direction and we can observe only a small angular portion
$\sim \gamma^{-1} \sim 0.01$ rad of the whole jet.
On the other hand, since the gravitational wave emission is anti-beamed,
it can be observed at larger viewing angles, so that we can see wider 
angular portion of the whole jet.
Therefore the information carried by the gravitational waves
is important to know the jet structure.
In this paper we investigate the gravitational wave 
from an inhomogeneous jet. We model the inhomogeneous jet in which
multiple sub-jets with a finite opening half-angle
are ejected to various directions.
The jet of GRBs should not be a single homogeneous
one since  a single explosion cannot produce a variable GRB.
We discuss what kind of information is carried by
the gravitational wave observed by DECIGO and BBO.
We also discuss the detectability of such events.

This paper is organized as follows.
In Sec.\ref{sec:memory}, we consider the gravitational wave
from a single jet acceleration.
We investigate the angular dependence of the gravitational
wave from a jet with a finite opening half-angle.
In Sec.\ref{sec:light},
we discuss the gravitational waveform from the GRB jet
adopting a unified model of the GRB proposed by some of authors
\cite{Yamazaki:2004ha}.
In Sec. \ref{sec:detect}, we calculate the strain amplitude
detected by the gravitational wave detector. We also
evaluate the maximum detectable distance to the GRBs
for such events by using DECIGO and BBO.
Finally, we summarize this paper and
discuss our results in Sec. \ref{sec:dis}.
We use the geometric units $G=c=1$ and the signature $(- + + +)$.

\section{Gravitational wave memory of a single jet}\label{sec:memory}
In this section we consider the gravitational waves
from a single jet acceleration.
Actually a single explosion cannot produce a variable GRB.
Many jets are to be ejected
since the total duration of the burst is usually much longer than the variability
timescale \cite{Kobayashi:1997}.
Many jets case will be discussed in the next section.

First let us remember the generic form of the gravitational wave memory
\cite{Braginsky,Thorne:1992}.
We consider that the system consists of freely moving masses 
before and after the burst.
Then the variation of the metric perturbation
(which is the gravitational wave memory) is given by
\begin{eqnarray}
\Delta h_{ij}^{\rm TT}=\Delta \sum_{A}
\frac{4 \gamma_{A} m_{A}}{r}
\left(\frac{\beta_{A}^{i} \beta_{A}^{k}}{1-\beta_{A} \cos \theta_{A}}
\right)^{\rm TT}.
\label{eq:memory}
\end{eqnarray}
Here $\Delta$ means the difference between before and after the burst,
$m_{A}$ is the mass labeled by $A(=1,2,\cdots)$,
$r$ is the distance from the source to the observer,
$\beta_{A}^{i}$ is the velocity of mass $A$,
$\gamma_{A}=(1-\beta_{A}^{2})^{-1/2}$ is the Lorentz factor of mass $A$,
and $\theta_{A}$ is the angle between $\beta_{A}^{i}$ and 
the direction from the source to the observer.
Small Latin indices ($i,j,\cdots$) run over spatial coordinates $x$, $y$, $z$, 
and the quantities $r$, $\beta_{A}^{i}$, $\theta_{A}$ are measured 
in the lab frame.
The TT part of $\Delta h_{ij}$ is obtained by
$\Delta h_{ij}^{\rm TT}=P_{ik} P_{jl} \Delta h_{kl}
-(1/2) P_{ij} (P_{kl} \Delta h_{kl})$
where $P_{ij}=\delta_{ij}-n_{i} n_{j}$ is the projection operator
and $n_{i}$ is the unit vector from the source to the observer
\cite{Gravitation:1973}.
The source quantities are to be evaluated at the retarded time.

The gravitational wave memory of a point mass acceleration is easily
calculated from equation (\ref{eq:memory})
\cite{Segalis:2001ns,Piran:2001ks,Piran:2002kw}.
Let us consider that a point mass $M$ is accelerated 
to a Lorentz factor $\gamma$.
Conservation of energy requires that the  mass of the accelerated particle
is $m=M/\gamma$.
We take a coordinate such that the $z$ axis is 
the direction from the source to the observer
and $\beta^{i} \propto (\sin\theta \cos\phi,\sin\theta \sin\phi,\cos\theta)$.
Then equation (\ref{eq:memory}) gives the amplitude of
the gravitational wave memory as
\begin{eqnarray}
\Delta h \equiv \Delta h_{+}+i \Delta h_{\times}=\frac{2\gamma m \beta^{2}}{r}
\frac{\sin^2\theta}{1-\beta\cos\theta} e^{2 i \phi},
\label{eq:point}
\end{eqnarray}
where $\Delta h_{+}\equiv \Delta h_{xx}^{\rm TT}=-\Delta h_{yy}^{\rm TT}$
and $\Delta h_{\times}\equiv \Delta h_{xy}^{\rm TT}=\Delta h_{yx}^{\rm TT}$
are two polarizations of the gravitational waves, plus and cross modes.
For $\gamma \gg 1$ and $\theta \ll 1$, we have
\begin{eqnarray}
|\Delta h| \simeq \frac{4\gamma m}{r}\frac{\theta^2}{\gamma^{-2}+\theta^2},
\end{eqnarray}
while for $\gamma \gg 1$ and $\theta \gg \gamma^{-1}$
\begin{eqnarray}
|\Delta h| \simeq \frac{2\gamma m}{r} (1+\cos\theta).
\end{eqnarray}
Therefore the gravitational wave memory is anti-beamed
with a similar amplitude
in almost all direction but little amplitude in 
the forward direction $\theta \alt \gamma^{-1}$.
The waveform has a form like a step function.
As we can see from equation (\ref{eq:point}),
the rise time of the amplitude from its initial to final value is 
about the time to reach $\beta \sim 1$ if the energy $\gamma m$ is conserved.
Thus the rise time is about $\delta t \sim 10^{7}$ cm$/c\sim 10^{-3}$ s
in the fireball model of the GRBs
\cite{Paczynski:1986px,Goodman:1986}.
If a counter-jet exists \cite{Yamazaki:2002wi},
we sum both contributions to obtain
\begin{eqnarray}
|\Delta h|=\frac{4\gamma m}{r}\sim 1.6 \times 10^{-22}
\left(\frac{2 \gamma m}{3\times 10^{51} {\rm erg}}\right)
\left(\frac{r}{1 {\rm Mpc}}\right)^{-1},
\label{eq:amp}
\end{eqnarray}
for $\gamma \gg 1$ and $\theta \gg \gamma^{-1}$.
(In this paper, however, we do not consider the counter jet
\footnote{From equation (\ref{eq:point}), we find that the ratio between
the amplitude of the gravitational wave due to the forward jet
and the one due to the counter jet is
$(1+\beta\cos\theta_v)/(1-\beta\cos\theta_v)$.
Since we consider the case of $\theta_v \alt 0.25$ in this paper,
this ratio is much larger than unity.
Therefore we can neglect the effect of the counter jet.}.)

In reality the opening half-angle of the GRB jet is 
larger than $\gamma^{-1}$ \cite{Frail:2001qp,Bloom:2003eq}.
Thus equation (\ref{eq:point}) is not appropriate for the GRB jet.
To examine the effects of the finite opening half-angle,
let us first consider a simple model, that is, 
an uniform thin jet with an opening half-angle $\Delta \theta$
that is accelerated to a Lorentz factor $\gamma$.
Since linear perturbation is additive,
the gravitational wave memory of the GRB jet is given by 
\begin{eqnarray}
\Delta h&=&\int \frac{\sin\theta d\theta d\phi}{\Delta \Omega}
H(\Delta \theta-|\theta-\theta_{v}|)
H\left[\cos\phi-
\left(\frac{\cos\Delta \theta-\cos\theta_{v}\cos\theta}
{\sin\theta_{v}\sin\theta}\right)\right]
\Delta h_{\bullet},
\label{eq:jet1}
\end{eqnarray}
where
$\Delta h_{\bullet}$ is the amplitude of a point mass 
in equation (\ref{eq:point}),
$H(x)$ is the Heaviside step function describing that the jet
is inside the cone,
$\theta_{v}$ is the viewing angle that the jet axis
(in the $xz$ plane) makes with the $z$ axis,
and $\Delta \Omega=\int_{0}^{2\pi} \int_{0}^{\Delta\theta} \sin\theta d\theta
d\phi=2\pi(1-\cos\Delta\theta)\simeq \pi(\Delta\theta)^2$
is the solid angle of the jet.
Replacing $\phi$ with $-\phi$ in equation (\ref{eq:jet1}),
we can show $\Delta h_{-}=0$.
For $\Delta h_{+}$ we can simplify equation (\ref{eq:jet1}) as
\begin{eqnarray}
\Delta h_{+}=\int_{\max[0,\theta_{v}-\Delta\theta]}^{\theta_{v}+\Delta\theta}
d\theta
\frac{2\gamma m \beta^{2}}{r \Delta\Omega}
\frac{\sin^{3}\theta \sin 2\Delta\phi}{1-\beta\cos\theta},
\label{eq:jet}
\end{eqnarray}
where 
\begin{eqnarray}
\Delta \phi=\left\{\begin{array}{l}
\pi \quad : \Delta \theta>\theta_v ~{\rm and}~ 
0<\theta \le \Delta \theta - \theta_v,\\
\cos^{-1}\left[
\frac{\cos \Delta \theta - \cos \theta_v \cos \theta}
{\sin \theta_v \sin \theta}\right]
\quad {\rm :others}.
\end{array}\right.
\end{eqnarray}
As in the point mass case, 
the gravitational waveform becomes like a step function, and
the rise time of the amplitude 
is about the time to reach $\beta\sim 1$ if the energy $\gamma m$ is
conserved.

\begin{figure}[t]
\includegraphics[width=10cm]{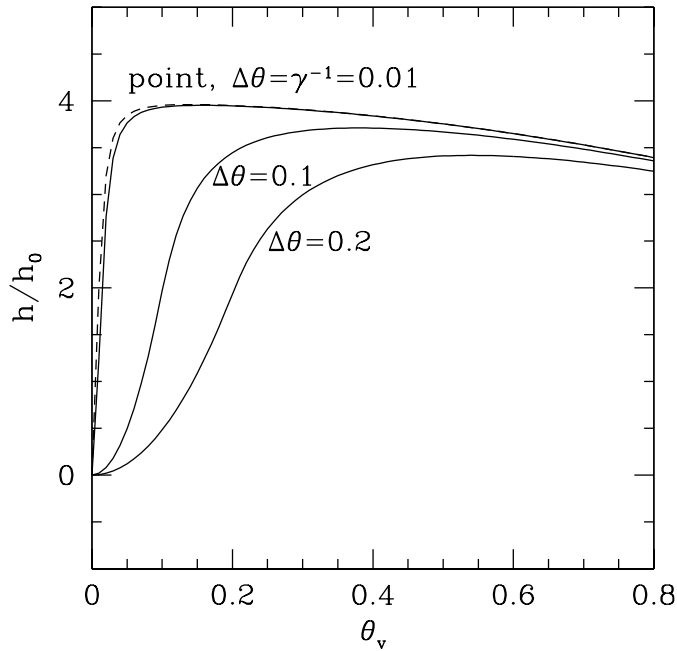}
\caption{
The amplitude of the gravitational wave memory
for the GRBs jet in equation (\ref{eq:jet})
is shown as a function of the viewing angle $\theta_{v}$ by solid
 lines. We adopt the Lorentz factor $\gamma=100$.
The opening half-angle of the jet is $\Delta\theta=0.01, 0.1$ and $0.2$
from top to bottom.
We can see that the amplitude is suppressed in the forward direction
$\theta_{v} \alt \Delta \theta$.
The amplitude for the point mass in equation
(\ref{eq:point}) is also shown by the dashed line.
Clearly the point mass approximation is good when 
the opening half-angle is smaller than the inverse of the Lorentz factor
$\Delta\theta < \gamma^{-1}$
or the viewing angle is larger than the opening half-angle
$\theta_{v} \agt \Delta \theta$.
The amplitude is normalized by
$h_0=\gamma m \beta^2/r \sim 2.7 \times 10^{-23} (\gamma m/10^{51} {\rm erg})
(r/1 {\rm Mpc})^{-1}$.}
\label{fig:angle}
\end{figure}

In Figure~\ref{fig:angle}, the amplitude of the gravitational wave memory 
for the GRB jet in equation (\ref{eq:jet})
is shown as a function of the viewing angle $\theta_{v}$ by solid lines.
We adopt the Lorentz factor $\gamma=100$.
The opening half-angle of the jet is $\Delta\theta=0.01, 0.1$ and $0.2$
from top to bottom.
We can see that, because of the anti-beaming effect,
the amplitude is suppressed in the forward direction,
that is, when the viewing angle is smaller than 
the opening half-angle of the jet $\theta_{v} \alt \Delta \theta$.
The amplitude of the memory for the point mass in equation
(\ref{eq:point}) is also shown by the dashed line.
We can find that the point mass is a reasonable approximation of the jet
when the opening half-angle is smaller than the inverse of the Lorentz factor
$\Delta\theta < \gamma^{-1}$
or the viewing angle is larger than the opening half-angle
$\theta_{v} \agt \Delta \theta$.
The opening half-angle
of the jet is usually small $\Delta \theta \sim 0.1$
\cite{Frail:2001qp,Bloom:2003eq} so that we observe the jet
from off-axis direction usually.
This means that the effects of the finite opening angle may not reduce the event rate of the gravitational waves from GRBs so much.

\section{Gravitational waveform from gamma-ray burst jets}\label{sec:light}

\subsection{Typical frequency}

The leading model of the GRB emission is the internal shock model
\cite{Kobayashi:1997}.
The internal shock occurs in the relativistic wind
when the fast moving flow catches up the slow one.
The wind can be modeled by a succession of relativistic shells.
A collision of two shells produces a single pulse light curve, whose
superposition makes a whole light curve.

When the GRB ejects many jets,
the gravitational waveform from the GRB differs from 
a single step function, but has many steps
since linear perturbation is additive.
There are three timescales that determine
the temporal structure of the gravitational wave memory of the GRB.
The first is the rise time of the gravitational wave amplitude due
to a single jet $\Delta T_{\rm rise}$.
As argued in the previous section,
$\Delta T_{\rm rise}$ is related to the acceleration time of the jet.
The second is the separation between successive jets $\Delta T_{\rm
sep}\sim 0.1$s.
The third is the total duration of the burst $\Delta T$,
which is about $\Delta T\sim 10$ s for long GRBs.
Then the typical order between three timescales is
\begin{eqnarray}
\Delta T_{{\rm rise}} \ll \Delta T_{{\rm sep}} \ll \Delta T.
\end{eqnarray}
Therefore the typical gravitational waveform should have
many steps. Each step corresponds to the ejection of a jet,
and the final amplitude of the gravitational wave memory is mainly
determined by the total energy of the jets ejected from the GRB.
The final amplitude is reached after the total duration $\Delta T$.
Thus the typical frequency in Fourier space of the gravitational waves is expected as 
\begin{eqnarray}
f_{c}\sim \frac{1}{\Delta T} \sim 0.1 \ {\rm Hz}. 
\end{eqnarray}
This band  is just  the decihertz band
of DECIGO and BBO.

\subsection{A unified model of GRBs}

Recently some of authors proposed a unified model of the GRBs \cite{Yamazaki:2004ha}.
In this model, the central engine of the long and short GRBs, XRFs
and X-ray rich GRBs is the same, and the apparent differences come
essentially from different viewing angles.
We assume that the jet of the GRB consists of multiple sub-jets.
These sub-jets have angular size of $\sim \gamma^{-1} \ll 0.1$
and are distributed within the whole GRB jet ($\sim 0.1$).
If many sub-jets point to our line of sight, the event looks like a long GRB,
while if a single sub-jet points to us, it looks like a short GRB.
If our line of sight is off-axis to any sub-jets, the event looks like
an XRF or X-ray rich GRB.

\begin{figure}[t]
\includegraphics[width=10cm]{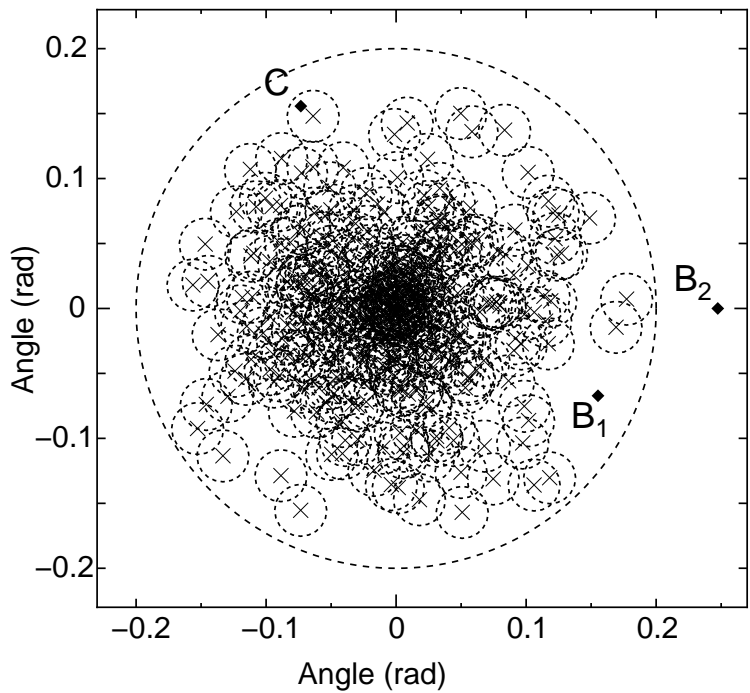}
\caption{
The angular distribution of $N_{\rm tot}=350$ sub-jets confined 
in the whole GRB jet in our simulation.
The whole jet has the opening half-angle of 
$\Delta\theta_{\rm tot}=0.2$~rad, while
the sub-jets have $\Delta\theta_{\rm sub}=0.02$~rad.
The axes and the angular size of sub-jets are represented by crosses
and the dotted circles, respectively.
 ``A'' represents the center of the whole jet and is hidden 
by the lines of sub-jets.
(Also refer to  Yamazaki, Ioka and Nakamura, Astrophys.J. 607 (2004) L103-L106
\cite{Yamazaki:2004ha}.)
}
\label{fig:dist}
\end{figure}

\begin{figure}[t]
\includegraphics[width=10cm]{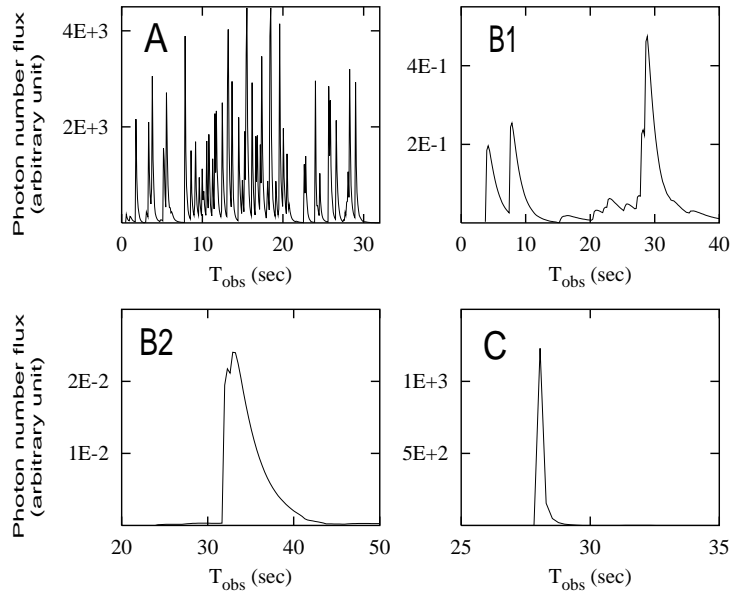}
\caption{
The observed X-ray and gamma-ray photon number flux from the multiple 
sub-jets,
corresponding the cases ``A''(the upper left),
``B$_1$''(the upper right), ``B$_2$''(the lower left) and
``C''(the lower right) in Figure~\ref{fig:dist}.
Here the flux in $30-400$ keV band for each case is shown.
(Also refer to Yamazaki, Ioka and Nakamura, Astrophys.J. 607 (2004) L103-L106
\cite{Yamazaki:2004ha}.)
}
\label{fig:gamma}
\end{figure}

An example of the sub-jet angular distribution is shown in Figure~\ref{fig:dist},
where $N_{\rm tot}=350$ sub-jets are ejected for $\Delta T=30$ sec
and each sub-jet has $\gamma=100$ and $\Delta \theta_{\rm sub}=0.02$ rad.
When we observe the source from the $\theta_v=0$ axis
(case~``A''), we see spiky temporal structures
(Fig. 3-A) and  may identify the case ``A'' as the long GRBs.
When the line of sight is away from any sub-jets
(cases~``B$_1$'' and ``B$_2$''),
soft and dim prompt emission, i.e.  XRFs
or X-ray rich GRBs are observed (Fig.~3-B$_{1}$, 3-B$_{2}$).
If the line of sight is inside an isolated sub-jet (case~``C''),
its observed pulse duration is much smaller than 
the case ``A'' (Fig.~3-C), so that we may observe the short GRB.
The details about the calculations of the X-ray and gamma-ray light curves
in Figure~\ref{fig:gamma} are given in \cite{Yamazaki:2004ha}.

\begin{figure}[t]
\includegraphics[width=10cm]{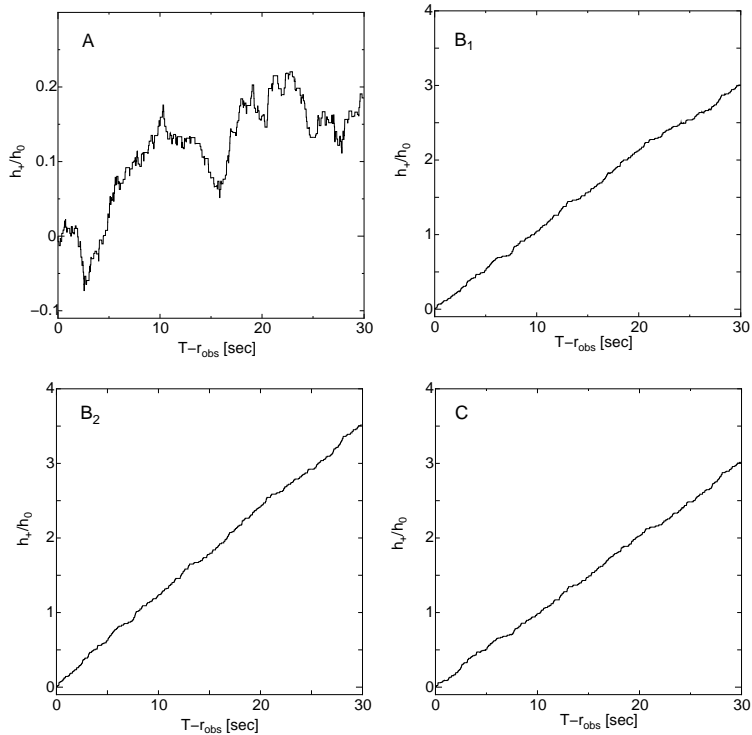}
\caption{
The waveform of the gravitational wave memory
from the GRB is shown.
There are three timescales that determine the temporal structure
of the waveform, the total duration $\Delta T\sim 30$ s,
the separation between successive jets
$\Delta T_{{\rm sep}}\sim \Delta T/N_{{\rm tot}}=0.086$ s
and the rise time of the amplitude due to a single jet
$\Delta T_{{\rm acc}}\sim 10^{-3}$ s
(although we neglect the rise time here).
The final amplitude is reached after the total duration $\Delta T$.
The normalization, $h_{0}$ is the same one in Figure \ref{fig:angle}.
}
\label{fig:light}
\end{figure}

In Figure~\ref{fig:light}, we show the plus mode of
the gravitational wave as a function of time assuming the same model as above.
Each panel (``A'', ``B$_1$'', ``B$_2$'' and ``C'')
corresponds to the viewing angle presented in Figure~\ref{fig:dist}
and the X-ray and gamma-ray light curves in Figure~\ref{fig:gamma}.
{}From Fig.1  for  $\Delta \theta\sim \gamma^{-1}$,
the point mass approximation is good.
Since $\Delta \theta_{\rm sub}=2 \gamma^{-1}$ in our case  
 we use the point mass approximation for each sub-jet
 and neglect the rise time of
the amplitude, for simplicity.  
Each step corresponds to the ejection of a sub-jet,
and the whole gravitational waveform has $N_{\rm tot}=350$ steps.
Since the sub-jets are ejected for $\Delta T=30$ sec, the final amplitude
is reached after the duration $\Delta T$.

We  find that the final amplitude in case ``A'' is
much smaller than the other cases
and goes up and down more frequently than the others.
This is because each memory of the sub-jet is proportional to $e^{2 i \phi}$
from equation (\ref{eq:point}) and the distribution of $\phi$ is
nearly isotropic for an on-axis observer like case ``A'' 
so that each memory tends to be canceled.
On the other hand, if the viewing angle is large,
the distribution of $\phi$ is not isotropic and 
either of two polarizations
dominates the other one. Therefore, the amplitudes
in cases ``${\rm B}_1$'',``${\rm B}_2$'' and ``C'' increase monotonously.
In addition, their waveforms resemble each other
because the gravitational wave  emitted by each sub-jet
is  nearly isotropic $\propto (1+\cos \theta)$ for $\theta \lesssim 0.2$.

Comparing the gravitational waveform
with the gamma-ray one in Figure~\ref{fig:gamma},
we can see that the gravitational wave amplitude is roughly
anti-correlated with the photon flux. In case ``A'', the gamma-ray is
strong while the gravitational wave is weak. In cases  ``${\rm B}_1$''
and ``${\rm B}_2$'', the gamma-ray is weak while the gravitational wave
is strong so that we may confirm the existence of many off-axis sub-jets
which are not identified by the gamma-ray. In case ``'C', the gamma-ray
is  short and strong while the gravitational wave is long and strong.
This means that the gravitational waves carry information on the off-axis jets.
Thus, the gravitational waves are indispensable to explore the
structure of the jet.
For instance, in other models of the XRFs, emission comes from on-axis jets,
so that the gravitational wave would have less amplitude.
In other words, the gravitational wave can  be used to verify 
or refute models of GRBs including our unified model of  GRBs.

\section{Detectable distance}\label{sec:detect}
\begin{figure}[t]
\includegraphics[width=10cm]{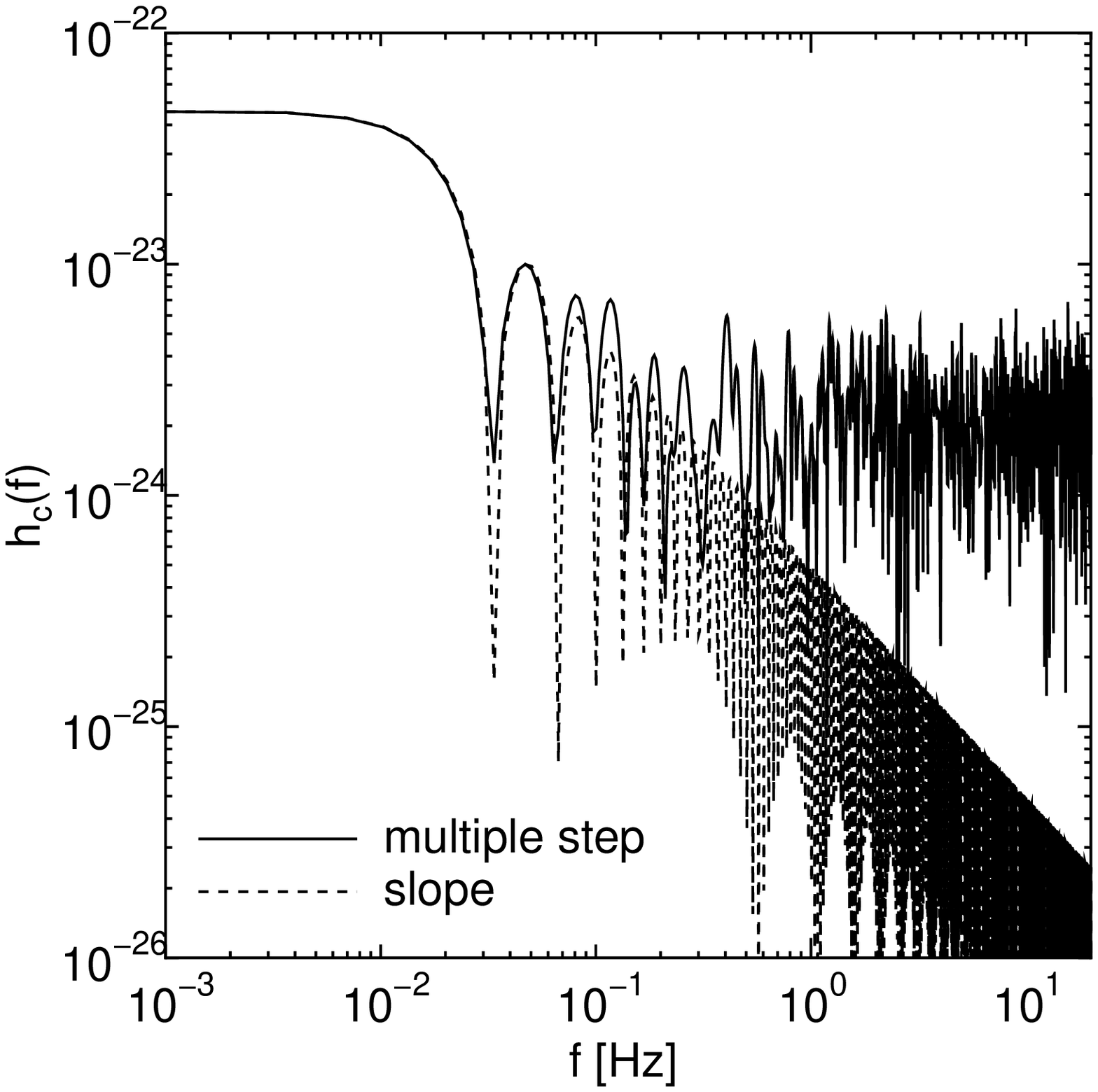}
\caption{
The characteristic amplitude $h_{c}$ of the gravitational wave memory
is plotted as a function of frequency for the whole GRB jets.
The solid line represents $h_c$ for the multiple step
waveform (case ``B$_2$'')
with the total energy $2 \gamma m = 3\times 10^{51}$ erg,
the duration $\Delta T=30$ sec
and the distance to the source $r=1$Mpc.
The dashed line represents $h_c$ for the slope waveform
with the duration $t_m=30$ sec.
Here the amplitude $\Delta h_m$ in equation (\ref{eq:slope}) is chosen so that
both amplitudes coincide
in the low-frequency limit.
}
\label{fig:char_amp}
\end{figure}

For the detector with a peak sensitivity $h_{peak}$ at frequency
$f_{peak}$, the optimal sensitivity to the gravitational wave memory
is $\sim h_{peak}$ as long as the rise time of the memory $t_{m}$
is smaller than $\sim 1/f_{peak}$ \cite{Braginsky,Thorne:1992}.
For example, we consider a simple waveform
as follows:
\begin{eqnarray}
h(t)=\left\{\begin{array}{ll}
0 & : t<0,\\
\Delta h_m (t/t_m) & : 0<t<t_m, \\
\Delta h_m & : t>t_m,
\end{array}\right.
\label{eq:slope}
\end{eqnarray}
which we call the slope waveform.
We can calculate the Fourier component of this waveform,
$\tilde h=\int_{-\infty}^{\infty} h(t) e^{i 2\pi f t} dt$,
which yields \footnote{
Note that $\int_{-\infty}^{\infty} e^{i 2 \pi f t} H(t) dt
=(1/2)[\delta (f)+ {\rm Pv} (i/\pi f)]$
where $\rm Pv$ denotes the principal value}
\begin{eqnarray}
|\tilde h|^{2}=\frac{(\Delta h_m)^{2}}{8\pi^{4} f^{4} t_{m}^{2}}
(1-\cos 2 \pi f t_{m}).
\label{eq:char-slope}
\end{eqnarray}
Note that $|\tilde h|^2 \simeq (\Delta h_m/2\pi f)^2$ for $f \ll 1/t_{m}$.
In Figure \ref{fig:char_amp}, 
we plot the characteristic amplitude of the gravitational wave
memory, $h_{c}(f)=2 f |\tilde h|$ for $t_m=30 \, {\rm sec}$
by the dashed line.
We find that $h_c(f)$ is constant at $f\alt 1/t_m$, while
it decreases for  $f\agt 1/t_m$.
Therefore the detector has the maximal sensitivity $h_{{\rm peak}}$
for the gravitational wave memory
as long as $1/t_m \alt f_{{\rm obs}}$.
We also plot $h_c(f)$ for the multiple step waveform (case ``B$_2$'')
by the solid line.
Here we fix the total energy and the distance of the source as
$2m\gamma=3\times 10^{51}$erg and $1$Mpc, respectively.
For this case, the amplitude is larger than that for
the slope waveform at $f \agt 1/\Delta T$.
This is because the waveform in the sub-jet model has
the smaller timescale, $\Delta T_{{\rm sep}}$.

\begin{figure}[t]
\includegraphics[width=10cm]{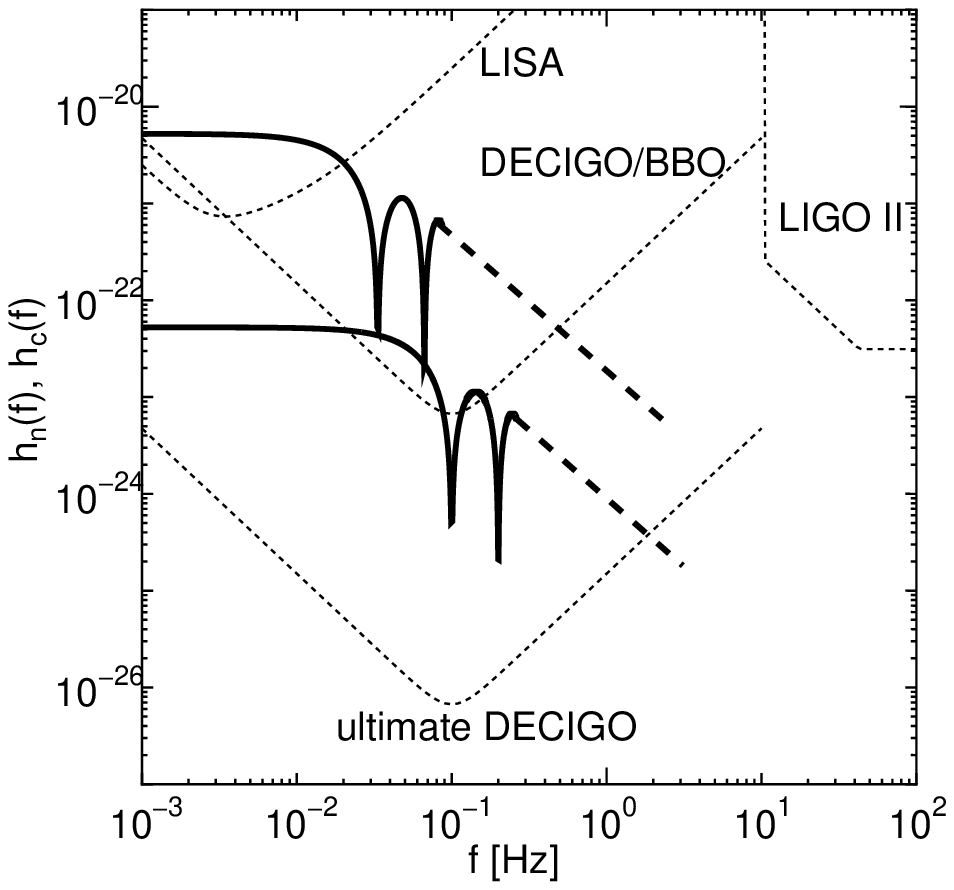}
\caption{
The noise amplitude $h_{n}$ of LIGO II, LISA and DECIGO/BBO is also plotted
by the dotted line.
The characteristic amplitude for the gravitational wave memory
of GRBs is also plotted by the bold solid line.
The upper bold solid line corresponds to a GRB with
the total energy $2 \gamma m = 3\times 10^{51}$erg, the duration
$\Delta T=30$sec at the $10$kpc, and the lower one to a GRB
with $2 \gamma m = 3\times 10^{51}$erg , $\Delta T=10$sec at the $1$Mpc.
The characteristic amplitude for the slope waveform
with $\Delta h_m=4m\gamma\beta^2/r$ is used
for the low-frequency region, $f\alt 1/\Delta T$.
The amplitude at $f\agt 1/\Delta T$ (plotted by the bold dashed line)
depends on the structure of the GRB jet.
The SNR can be obtained by equation (\ref{eq:snr}).
}
\label{fig:spec}
\end{figure}

In Figure~\ref{fig:spec}, we plot the noise amplitude
$h_{n}(f)=[5 f S_{h}(f)]^{1/2}$
for LIGO II, LISA and DECIGO/BBO\footnote{
The LIGO II noise curve is taken from \cite{Flanagan:1997sx}.
We adopt 
$S_{h}(f)=1.22 \times 10^{-51} f^{-4}+
2.11 \times 10^{-41} + 1.22 \times 10^{-37} f^{2}$ Hz$^{-1}$
for LISA \cite{Finn:2000sy},
and $S_{h}(f)=4.5 \times 10^{-51} f^{-4}+ 4.5 \times 10^{-45} f^{2}$
Hz$^{-1}$ for DECIGO/BBO \cite{Seto:2001qf},
where $f$ is in unit of Hz.},
where $S_{h}(f)$ is the spectral density of the  detector's strain noise 
at frequency $f$, and $5$ comes from sky-averaging
\cite{Thorne:1987}.
We also plot the characteristic amplitude of
the gravitational wave memory by the bold line.
For the low-frequency region, $f\alt 1/\Delta T$,
we use the result for the slope waveform
with $\Delta h_m=4m\gamma\beta^2/r$,
while we plot it by the bold dashed line
in the high-frequency region, $f\agt 1/\Delta T$,
because it depends on the structure of the GRB jet.
The upper bold line corresponds to a GRB with
$(\Delta T, r, 2\gamma m)
=(30{\rm sec}, 10{\rm kpc}, 3\times 10^{51}{\rm erg})$,
and the lower one to a GRB with
$(\Delta T, r, 2\gamma m)
=(10{\rm sec}, 1{\rm Mpc}, 3\times 10^{51}{\rm erg})$.
The signal-to-noise ratio (SNR) is given by
\begin{eqnarray}
{\rm SNR}^{2}=\int_0^{\infty}
d(\ln f) \frac{h_{c}(f)^{2}}{h_{n}(f)^{2}},
\label{eq:snr}
\end{eqnarray}
thereby one can roughly read off the SNR from Figure~\ref{fig:spec}.
When the duration of GRBs is shorter than $10$sec,
the detectable distance reaches the Local Group ($\sim 1$Mpc)
by using the practical DECIGO and BBO.
On the other hand, the detectable distance is reduced 
when the duration is longer than $10$sec.
If a GRB happens in our galaxy, we can get the information on the
high-frequency region $f\agt 1/\Delta T$.
This may make it possible for us not only to detect the gravitational
wave memory, but also to extract important information about the structure
of the GRB jet from the observed gravitational wave.

\section{Summary and Discussion}\label{sec:dis}
In this paper, we have investigated the gravitational wave emitted
in the acceleration phase of the GRB jet.
At first, we have considered the angular distribution of
a GRB jet with a finite opening half-angle $\Delta\theta$.
As a result, we have shown that we can deal with the jet
as a point mass if the opening half-angle is smaller
than the inverse of the Lorentz factor, $\gamma^{-1}$,
or the viewing angle $\theta_v$ is larger than the opening half-angle.
We have also shown that the gravitational radiation is
weak in the angular size $\sim \gamma^{-1}$.
Next, we have considered the waveform of the gravitational wave from the
GRBs. We have found that many steps usually appear in the waveform.
This has been explicitly shown for the multiple sub-jet model of GRBs.
If we observe the GRBs from the front, the observed amplitude
of the gravitational wave is suppressed because the wave from each sub-jet
tends to cancel one another.
On the other hand, the amplitude increases monotonously
for the off-axis observer.
In addition, we have considered the detectability of GRBs
by using the practical DECIGO and BBO.
We have found that we may detect the gravitational wave emitted
from the GRB jet if it locates within $1$ Mpc.

Since the gravitational wave memory is anti-beamed,
almost all GRBs detected by the gravitational waves will
be observed with off-axis viewing angle.
The photon emission from such off-axis GRBs is
softer and dimmer than the on-axis ones due to the relativistic beaming
\cite{Ioka:2001,Yamazaki:2004pv}.
Therefore the GRBs detected by the gravitational wave memory
may be observed as XRFs \cite{Yamazaki:2002yb,Yamazaki:2002sy}
or rather softer events in the UV/optical bands \cite{Yamazaki:2002wi}.
The prompt electromagnetic emission is too dim to be observed
in the case of large viewing angle. Even then,
bright radio afterglows can be seen several months after the GRB explosion
since the relativistic beaming effect becomes weak
\cite{Levinson:2002aw,Totani:2002ay}.
The association with a bright radio source might help us to identify
the signal of the gravitational wave memory.
In addition, the association with supernovae observed in the optical band 
may be also helpful.
XRFs may be GRBs with large viewing angles
\cite{Yamazaki:2004pv,Yamazaki:2004ha}.
The gravitational waves will be observed prior to the XRFs
with the time delay being longer for softer and dimmer XRFs
\cite{Ioka:2001}.
Thus the gravitational waves allow one to test the off-axis
GRB model for the XRFs.

Unfortunately the event rate of the gravitational waves associated
with the detectable GRBs may be small.
Since the true GRB rate is estimated as $\sim 250$ Gpc$^{-3}$ yr$^{-1}$
\cite{Frail:2001qp},
the event rate within the detectable distance $\sim 1$ Mpc
(see \S~\ref{sec:detect}) is only $3\times 10^{-7}$ yr$^{-1}$,
that is very small compared with the detection rate of the gravitational
wave by the central engine \cite{vanPutten:2002ui}.
However the gravitational wave event rate without detectable GRBs
could be much higher. Recently, many observations support
a massive stellar origin for the long GRB
\cite{Meszaros:2001vi,Zhang:2003uk} and
it is widely assumed that a massive collapsing star
(``collapsar'') is the progenitor
\cite{Woosley:wj,Paczynski:1997yg}.
Suppose the GRB jets have the same total energy.
Although the GRB jet cannot break through the star
because of weak collimation or it is baryon-loaded,
so that $2 \alt \gamma \alt 30$,
the resultant gravitational wave may have
same amplitude as usual succeeded GRBs.
Hence the detection rate of such events may be improved.
Furthermore we calculate the detection rate for mean events.
However the total energy is distributed around these mean values.
Then the detections will be dominated by rare, energetic events.
For instance, even if only $\sim 2\%$ events are 10 times more energetic,
the detection rate will be increased by $\sim 2\% \times 10^{3} \sim 20$,
since we can observe 10 times farther events.
Even if we take account of the above, the detection rate is
$6\times 10^{-3}$ yr$^{-1}$ optimistically.
We need $\sim 10$ times better sensitivity than that of practical
DECIGO/BBO for the detection rate $\sim 1$ yr$^{-1}$.

The gravitational wave memory does not depend on the form of energy.
For example, even if the jet starts out as a pure Poynting flux jet
\cite{Lyutikov:2003ih,Levinson:2003je},
the memory is observed in the similar way.
One may also think that the memory is lost if the jet is
decelerated, for example, at the internal shock, the external
shock and the termination shock inside the massive star.
However at the internal and external shock
almost all kinetic energy of the jet goes into the photon emission
whose memory also remains.
(In equation (\ref{eq:memory}), $\gamma m$ should be interpreted
as the photon energy, and the photon speed $\beta$ must be set to $1$.)
Thus the amplitude of the gravitational wave memory does not change so much.
At the termination shock between the jet and the massive stellar envelope,
the kinetic energy may be converted to the neutrino emission
\cite{Meszaros:2001ms,Razzaque:2003uw}.
Since the neutrino also leave the memory,
the gravitational wave memory may not be reduced so much as long as
the cooling is not severe in the jet head \citep{Razzaque:2003uw}.

\begin{acknowledgments}
We are grateful N.~Seto, K.~Kohri, K.~Asano and R.~Takahashi 
for useful comments.
This work was supported in part by the Grant-in-Aid for the 21st Century COE
``Center for Diversity and Universality in Physics''
and the Monbukagaku-sho Grant-in-Aid, 
No.660 (KI), No.14047212 (TN), No.14204024 (TN) and No.05008 (RY),
and by the Center for Gravitational Wave Physics under
the National Science Foundation cooperative agreement PHY 01-14375 (KI).
\end{acknowledgments}


\end{document}